\documentclass[conference]{IEEEtran}
\IEEEoverridecommandlockouts
\usepackage{cite}
\usepackage{amsmath,amssymb,amsfonts}
\usepackage{algorithmic}
\usepackage{graphicx}
\usepackage{textcomp}
\usepackage{xcolor}
\usepackage{hyperref}
\usepackage{booktabs}
\usepackage{multirow}

\def\BibTeX{{\rm B\kern-.05em{\sc i\kern-.025em b}\kern-.08em
    T\kern-.1667em\lower.7ex\hbox{E}\kern-.125emX}}

\begin{document}

\title{Cybersecurity in Edge Computing: A Trust-Aware Federated Hybrid Intrusion Detection Framework}

\author{
\IEEEauthorblockN{Zawad Yalmie Sazid\textsuperscript{1}, Robert Abbas\textsuperscript{2}}
\IEEEauthorblockA{\textsuperscript{1,2}Victoria University, Sydney, Australia \\
Email: zawad.sazid@live.vu.edu.au, robert.abbas@vu.edu.au}
}

\maketitle

\begin{abstract}
Edge computing has emerged as one of the most critical computing paradigms in modern distributed systems by migrating data processing closer to the end users and Internet of Things (IoT) devices. While this paradigm decentralizes processes, minimizes latency, and reduces backhaul bandwidth congestion, it exponentially enlarges the cyber-attack surface. Heterogeneous, resource-constrained edge devices deployed across unmanaged administrative domains present highly vulnerable targets. To address these vulnerabilities without compromising strict global data privacy regulations, this paper proposes a novel Trust-Aware Federated Hybrid Intrusion Detection Framework (TA-FHIDF). The proposed framework integrates an Autoencoder, a One-Dimensional Convolutional Neural Network (1D-CNN), and a Bidirectional Long Short-Term Memory (BiLSTM) model into a unified, localized deep learning engine capable of autonomous spatial and temporal feature extraction. To maintain strict data privacy, the model is trained collaboratively via federated learning, ensuring that raw network telemetry remains isolated at the local gateway. Furthermore, to defend against adversarial attacks where a physically compromised node might submit corrupted weights (model poisoning), we introduce a robust server-side trust-aware aggregation mechanism. This mechanism mathematically evaluates client reliability using a cosine similarity metric before any global model integration occurs. Comprehensive empirical evaluations against traditional machine learning baselines across multi-vector cybersecurity benchmark datasets (UNSW-NB15, CICIDS2017, and Edge-IIoTset) demonstrate the framework's superior detection accuracy, rapid convergence, and high Byzantine fault tolerance under adversarial attack scenarios.
\end{abstract}

\begin{IEEEkeywords}
Edge Computing, Federated Learning, Intrusion Detection, Hybrid Deep Learning, Cybersecurity, Trust-Aware Aggregation, Byzantine Fault Tolerance.
\end{IEEEkeywords}

\section{Introduction}
\subsection{Research Background}
Edge computing has fundamentally transformed the architecture of modern distributed systems by shifting computational workloads from centralized cloud infrastructures directly to the network edge, closer to end-users and Internet of Things (IoT) devices \cite{shi2016, satyanarayanan2017}. This paradigm shift is primarily driven by the necessity to reduce network latency and optimize bandwidth consumption, which are critical requirements for real-time, mission-critical applications in domains such as smart manufacturing, autonomous systems, and digital healthcare \cite{shi2016}. The rapid deployment of 5G-Advanced networks has further accelerated this transition, demanding immediate, localized data processing capabilities.

However, while decentralization optimizes performance, it simultaneously creates a massive and highly distributed cyber-attack surface. The proliferation of heterogeneous, resource-constrained edge devices across unmanaged or physically accessible administrative domains makes them highly susceptible to localized exploitation \cite{satyanarayanan2017, zhukabayeva2025}. Consequently, protecting these decentralized edge environments against sophisticated, multi-stage cyber threats has emerged as a paramount challenge in contemporary distributed computing research. To counter these vulnerabilities, the industry is increasingly migrating toward deploying hybrid deep learning anomaly detection models directly on edge devices, thereby eliminating the need to centralize sensitive data pools \cite{altunay2023, baidar2025}. 

\subsection{Existing Challenges}
Designing a robust intrusion detection framework for edge environments is complex due to competing operational requirements. Traditional security models rely on complete network visibility, necessitating the continuous export of all local telemetry to a central cloud server. This approach is fundamentally incompatible with next-generation edge designs, which require strict decentralization to safeguard proprietary telemetry, minimize response latency, and comply with global privacy regulations (e.g., GDPR, HIPAA). Attempting to bridge this gap via federated and hybrid deep learning introduces distinct computational, networking, and cryptographic challenges:

\begin{itemize}
    \item \textbf{Data Privacy and Regulatory Compliance:} Centralizing raw network telemetry exposes sensitive operational topologies. Next-generation frameworks must ensure that data remains strictly localized, utilizing cryptographic parameter exchange rather than plain payload transmission.
    \item \textbf{Data Heterogeneity (Non-IID):} Edge gateways process highly localized traffic. Consequently, standard federated aggregation techniques (like FedAvg) often suffer from severe weight divergence because the data distributions across nodes are non-independently and identically distributed (non-IID).
    \item \textbf{Hardware Limitations:} Edge devices lack the CPU, memory, and thermal budgets of cloud data centers. Advanced deep neural networks must be mathematically compressed and optimized to allow for real-time localized threat inference without exhausting edge hardware.
\end{itemize}

\subsection{Research Problem and Objectives}
Driven by these inherent vulnerabilities, the overarching research problem is that current distributed intrusion detection systems fail to simultaneously balance strict data privacy, high detection accuracy across multiple vectors, algorithmic robustness against adversarial client poisoning, and feasibility on constrained hardware.

To systematically address this, this research pursues the following core objectives:
\begin{enumerate}
    \item \textbf{RO1:} Design a localized hybrid deep learning intrusion detection architecture (Autoencoder, 1D-CNN, BiLSTM) to automate high-dimensional spatial and temporal feature extraction on constrained edge nodes.
    \item \textbf{RO2:} Develop a privacy-preserving collaborative learning framework that eliminates the need to backhaul raw telemetry.
    \item \textbf{RO3:} Design and integrate a server-side trust-aware aggregation mechanism that dynamically evaluates and filters out heterogeneous or malicious parameter updates using cosine similarity metrics.
    \item \textbf{RO4:} Conduct a rigorous comparative ablation study evaluating detection accuracy and false-positive reduction against traditional machine learning baselines across multiple benchmark datasets.
\end{enumerate}

\section{Literature Review}
To provide a solid theoretical and empirical basis for the proposed architecture, the existing literature is systematically classified into four critical research domains.

\subsection{Foundations of Edge Computing and Network Security}
The transition from cloud-centric networks to decentralized edge architectures has been heavily documented, as has the corresponding rise in cybersecurity vulnerabilities. Shi et al. \cite{shi2016} formalized the foundational vision of edge computing as a necessary paradigm shift to address severe network latency and backhaul bandwidth saturation caused by exponential IoT growth. While they successfully established the structural baseline, their framework remained largely theoretical regarding active security mechanisms. Satyanarayanan \cite{satyanarayanan2017} expanded on this by highlighting the critical security vulnerabilities associated with deploying heterogeneous, resource-constrained edge nodes in unmanaged domains. 

In early attempts to secure IoT structures, Dong and Wang \cite{dong2016} applied centralized deep learning Intrusion Detection System (IDS) models. While successful in proving that automated feature extraction outperforms rule-based firewalls, their centralized methodologies fail in modern edge settings because transferring massive volumes of raw IoT data to a central cloud server introduces prohibitive latency. Exploring decentralized alternatives, Niranjanamurthy et al. \cite{niranjanamurthy2019} evaluated blockchain technology. However, they concluded that consensus protocols introduce prohibitively high computational overhead, rendering them unsuitable for lightweight edge threat mitigation. Furthermore, Saba et al. \cite{saba2022} established the necessity of behavior-based anomaly detection in digital healthcare but were limited by static, outdated datasets unable to account for real-time concept drift. This literature highlights a clear gap: the need for a mathematically robust, decentralized machine learning defense mechanism directly at the physical edge layer.

\subsection{Deep Learning and Hybrid Architectures for IDS}
To extract complex spatial and temporal features from network traffic, researchers have increasingly applied deep learning. Mahadevappa et al. \cite{mahadevappa2024} successfully deployed traditional machine learning algorithms (SVM, Decision Trees) at the edge for low-resource anomaly detection; however, these algorithms rely entirely on manual feature engineering and fail to capture complex, multi-stage zero-day attacks. Conversely, Nandanwar and Katarya \cite{nandanwar2024} developed \textit{AttackNet} for Industrial IoT (IIoT), demonstrating high efficacy but with an exceedingly narrow focus on singular attack vectors like DDoS.

To capture sophisticated threats, hybrid architectures are required. Albalawi et al. \cite{albalawi2023} designed a hybrid RNN-BiLSTM approach for edge-cloud deployments, drastically improving the detection of prolonged, time-delayed intrusions. However, their model lacked pre-processing spatial feature compression, meaning it easily overwhelmed limited edge memory during sequence unfolding. Altunay and Albayrak \cite{altunay2023} proposed a highly accurate CNN+LSTM model that eliminated manual feature engineering, but their strict reliance on centralized training fundamentally compromises data privacy. Most recently, Zhukabayeva et al. \cite{zhukabayeva2025} integrated neural networks directly at the IIoT edge for real-time inference, yet their system still relied on local raw data pooling, completely omitting collaborative mechanisms. Our research addresses this by porting a unified AE-CNN-BiLSTM architecture into a privacy-preserving federated environment, explicitly utilizing the Autoencoder for the vital spatial compression required by 5G-Advanced edge nodes.

\subsection{Privacy-Preserving Collaborative Learning}
Federated Learning (FL) enables distributed collaborative learning while preserving data privacy. McMahan et al. \cite{mcmahan2017} introduced the foundational Federated Averaging (FedAvg) algorithm, completely eliminating the need for raw telemetry transmission. However, standard FedAvg possesses a critical vulnerability: it relies on the unrealistic assumption that all participating clients are trustworthy, offering zero algorithmic defense against compromised nodes attempting data poisoning in unmanaged edge environments.

Evaluating federated systems requires robust datasets. While Moustafa and Slay \cite{moustafa2015} (UNSW-NB15) and Sharafaldin et al. \cite{sharafaldin2018} (CICIDS2017) provided excellent centralized benchmarks, Ferrag et al. \cite{ferrag2022} directly addressed federated constraints by creating the Edge-IIoTset, tailored explicitly for distributed learning. Utilizing such environments, Davies et al. \cite{davies2023} and Pecherle \cite{pecherle2025} empirically proved that localized class imbalances (non-IID data) severely distort local gradients, leading to catastrophic performance degradation in standard FedAvg. Chen et al. \cite{chen2026} attempted to solve evolving IoT threats with incremental federated learning, but their approach suffered from high latency and lacked server-side cryptographic validation. Standard FL's inherent trust in all clients necessitates a departure toward robust, secure aggregation.

\subsection{Trust-Aware and Resilient Aggregation Mechanisms}
To defend against poisoning and adversarial attacks, recent studies have explored mathematically evaluating client reliability before global model aggregation. Wang et al. \cite{wang2024} designed lightweight trust mechanisms based on historical reporting frequency, but this heuristic approach suffered lower accuracy against sophisticated zero-day attacks. Will and Kate \cite{will2024} proposed \textit{ZeroTrust-SecFL-IoT}, focusing on verifiable device identities. However, their identity perimeter approach meant a compromised but authenticated insider device could still easily submit poisoned weights undetected.

Li \cite{li2022} successfully implemented reputation-based trust frameworks, but evaluated them almost entirely on shallow machine learning models, leaving scalability with heavy deep neural networks unverified. Baidar et al. \cite{baidar2025} introduced a highly resilient, Byzantine-robust aggregation framework for 5G edge networks, demonstrating high convergence resilience. Mrabet \cite{mrabet2025} proposed \textit{TrustFed-CTI} for cyber threat intelligence sharing. The primary gap remaining is that advanced mathematical trust evaluations (e.g., deep cosine similarity gradient inspection) have rarely been integrated tightly with the complex hybrid deep learning models (AE-CNN-BiLSTM) required to process raw, multi-vector attack payloads directly at physical gateways. Our proposed framework actively bridges this gap.

\section{Methodology and System Architecture}
The proposed Trust-Aware Federated Hybrid Intrusion Detection Framework (TA-FHIDF) is designed to operate within strict hardware constraints, preserve data privacy, and mathematically defend against malicious federation updates. The framework decentralizes the threat detection process across three interconnected layers.

\subsection{Data Ingestion and Preprocessing Pipeline}
Raw network telemetry generated by IoT sensors is inherently noisy, high-dimensional, and prone to extreme statistical skew. Before deep learning ingestion, local Edge Nodes execute a rigorous preprocessing pipeline:
\begin{enumerate}
    \item \textbf{Cleaning \& Imputation:} Duplicate flows and zero-variance columns are removed. Missing network packet features undergo median substitution to prevent extreme volumetric outliers (e.g., massive DDoS payloads) from skewing the underlying statistical distribution.
    \item \textbf{Encoding \& Scaling:} Categorical protocol features (connection states, service types) are one-hot encoded. Continuous numerical features are standardized using Min-Max scaling to a $[0, 1]$ range, preventing volumetric domination during backpropagation.
    \item \textbf{Tensor Transformation:} To enable temporal sequence learning, static 2D tabular data is reshaped into 3D sequential tensors (Samples $\times$ Timesteps $\times$ Features) using overlapping sliding windows.
\end{enumerate}

\subsection{Local Edge Node Architecture (AE-CNN-BiLSTM)}
To automate spatial-temporal feature extraction, the framework deploys a multi-stage topology directly at the edge gateway. Let $x \in \mathbb{R}^d$ denote the preprocessed, high-dimensional network features.

\textbf{1) Dimensionality Reduction via Autoencoder (AE):}
To prevent memory saturation on constrained edge hardware, an Autoencoder compresses the input vector $x$ into a highly compact latent representation $z$:
\begin{equation}
z = \sigma(W_e x + b_e)
\end{equation}
Where $W_e$ is the encoder weight matrix, $b_e$ is the bias, and $\sigma$ is the non-linear activation function (ReLU).

\textbf{2) Spatial Feature Extraction via 1D-CNN:}
The compressed features $z$ are passed to a 1D-Convolutional Neural Network to extract localized spatial correlations indicative of payload anomalies:
\begin{equation}
C_i = \text{ReLU}(W_c * z_{i:i+k-1} + b_c)
\end{equation}
Where $*$ denotes the convolution operation and $k$ represents the kernel size.

\textbf{3) Temporal Sequence Tracking via BiLSTM:}
To capture the temporal sequence of multi-stage attacks over time, the CNN outputs are fed into a Bidirectional Long Short-Term Memory layer. The BiLSTM computes forward $\overrightarrow{h_t}$ and backward $\overleftarrow{h_t}$ hidden states, forming the final temporal representation before Softmax classification:
\begin{equation}
h_t = \overrightarrow{h_t} \oplus \overleftarrow{h_t}
\end{equation}

\subsection{Privacy-Preserving Federated Training}
The framework completely decentralizes the collaborative learning process. Edge nodes execute local backpropagation strictly on their isolated datasets. Raw network telemetry never crosses external administrative boundaries. Only cryptographically secured model parameter updates ($\Delta w$) are exchanged between the distributed edge gateways and the central cloud coordinator, effectively eliminating backhaul bandwidth congestion and strictly adhering to global privacy mandates.

\subsection{Server-Side Trust-Aware Aggregation}
To secure the federated topology against adversarial compromise (e.g., label flipping, gradient corruption), the central server operates a dynamic Trust Management Layer. Standard FedAvg is bypassed. Instead, before global model aggregation occurs, the central server mathematically evaluates the reliability of each client update $\Delta w_i^t$ at communication round $t$.

The server calculates the cosine similarity between client $i$'s update and the global average update to generate a dynamic trust score $T_i$:
\begin{equation}
T_i = \frac{\Delta w_i^t \cdot \Delta w_{global}^t}{||\Delta w_i^t|| \ ||\Delta w_{global}^t||}
\end{equation}

The global model is then aggregated using a weighted average based exclusively on these computed trust scores:
\begin{equation}
w^{t+1} = w^t + \frac{\sum_{i=1}^K T_i}{\sum_{j=1}^K T_j} \Delta w_i^t
\end{equation}
This algorithmic intervention dynamically isolates and throttles malicious or heavily non-IID updates, mathematically ensuring Byzantine resilience and stable global convergence.

\begin{figure*}[htbp]
    \centering
    \includegraphics[width=0.85\textwidth]{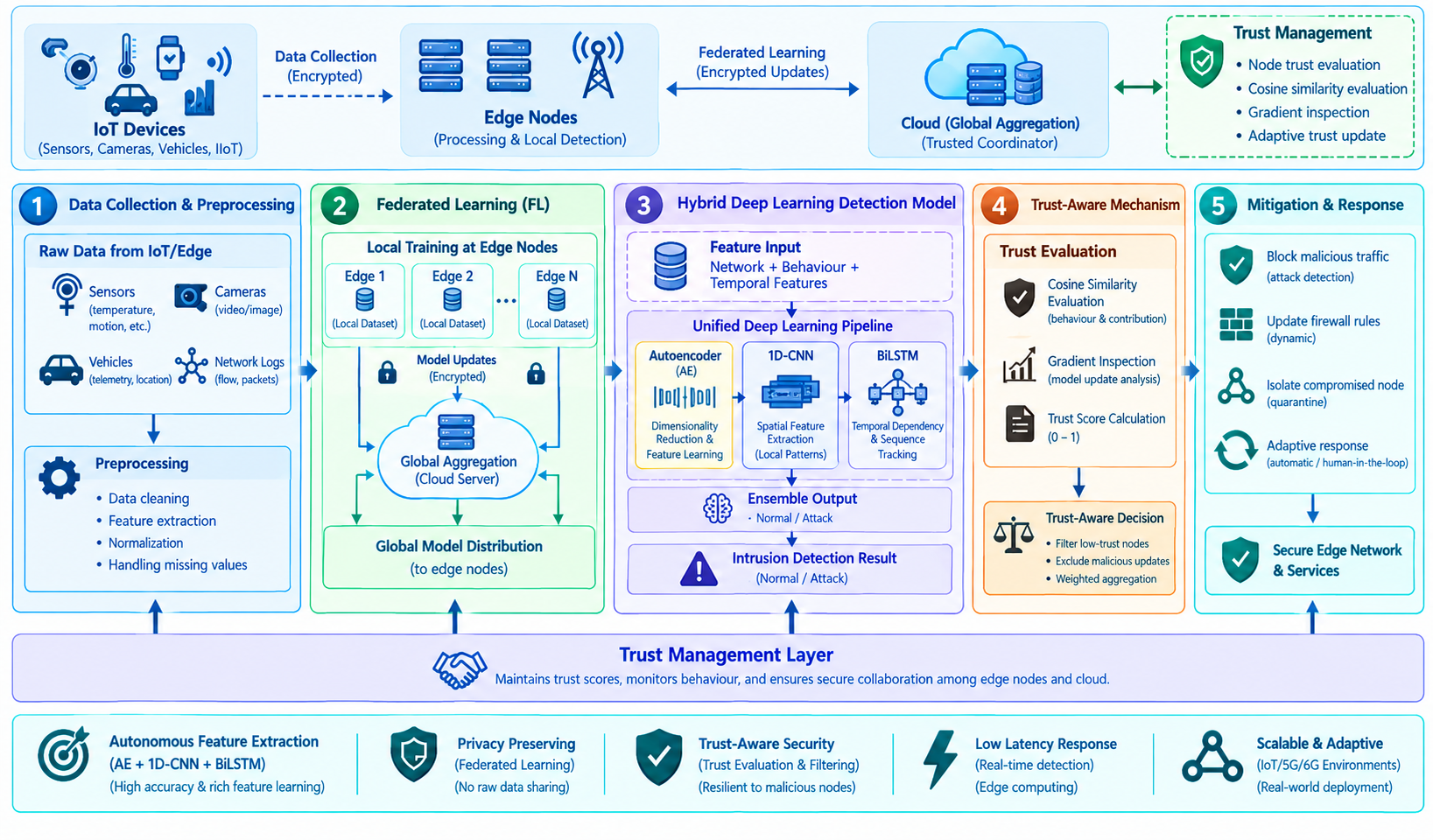} 
    \caption{Proposed Trust-Aware Federated Hybrid Intrusion Detection Framework (TA-FHIDF) Architecture}
    \label{fig:model_framework}
\end{figure*}

\subsection{Proposed Architectural Framework}
The primary challenges of edge computing systems that we can divide into three parts: the limitation of hardware resources, stringent privacy regulations, and continued threats from a compromised node. This paper proposes a unified architecture of the system to tackle these problems. It includes a new method of hybrid intrusion detection framework that fuses trust-aware decentralized teamwork across the network taking benefits from local deep learning.

Performing real-time threat detection at the edge without ever sending raw network traffic to a central cloud preserves sensitive IoT data in its entirety and completely outside of ownership by a third party. Moreover, a powerful trust-evaluation mechanism is embedded directly into the central server to avoid malicious insider attacks from nodes that attempt to poison the model.

As illustrated in Fig. \ref{fig:model_framework}, the visual layout of this architecture breaks down the data's journey across five connected stages, starting from the collection of raw network traffic and ending with a secure, globally updated model.

\subsubsection{Architecture Overview and Data Ingestion}
The proposed architectural framework initiates at the physical environment, where heterogeneous IoT and IIoT devices generate continuous network traffic. To prepare this raw telemetry for deep learning ingestion, the localized Edge Nodes execute a rigorous preprocessing pipeline.

\subsubsection{Unified Hybrid Deep Learning Engine}
Following preprocessing, the localized network traffic is processed by the core detection engine deployed directly at the edge gateway. This tripartite pipeline fully automates feature extraction independent of manual engineering.

\subsubsection{Decentralized Federated Learning Pipeline}
To satisfy strict data privacy regulations and optimize communication efficiency, the framework completely decentralizes the collaborative learning process. Edge nodes execute local backpropagation strictly on their isolated datasets, ensuring raw network telemetry never crosses external administrative boundaries.

\textbf{Trust-Aware Aggregation and Mitigation}
To secure the federated topology against adversarial compromise, the central cloud server operates a dynamic Trust Management Layer. The server mathematically evaluates every individual client update against the global consensus using cosine similarity gradient inspection. Upon anomaly detection, the localized mitigation layer immediately blocks malicious traffic, updates firewall rules dynamically, and quarantines compromised endpoints to maintain secure edge services.

\begin{figure*}[htbp]
    \centering
    \includegraphics[width=0.85\textwidth]{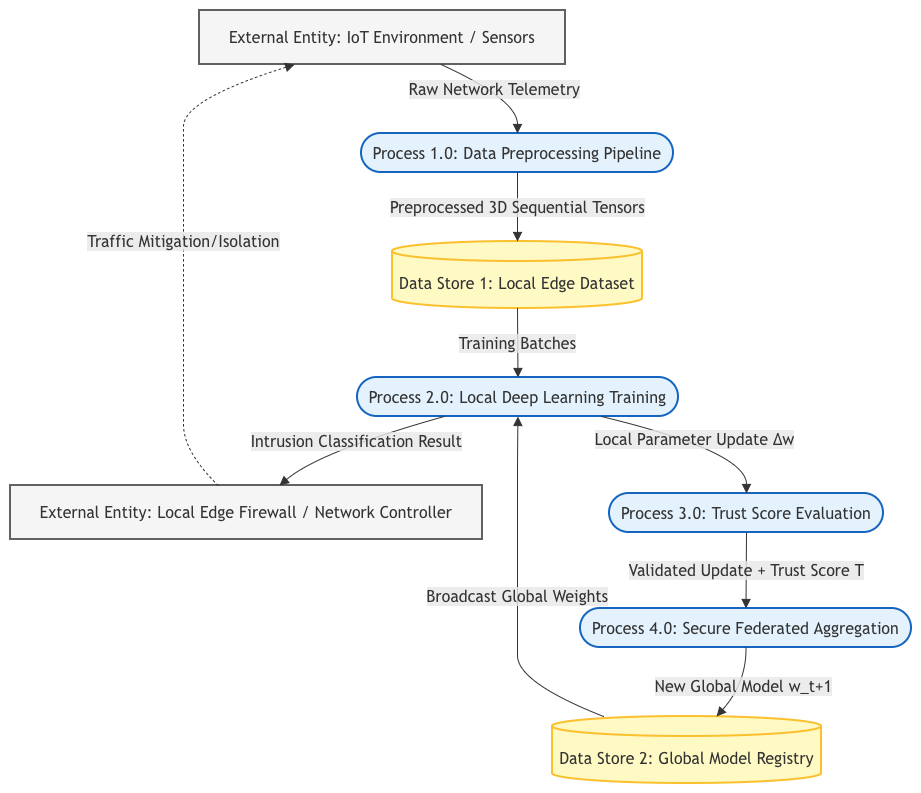} 
    \caption{Data Flow Diagram illustrating secure telemetry handling and cryptographic parameter exchange}
    \label{fig:dataflow_diagram}
\end{figure*}

\subsection{Dataflow Diagram for the Model}
To fully map the movement, transformation, and storage of information within the proposed architecture, a Data Flow Diagram (DFD) was constructed. As illustrated in Fig. \ref{fig:dataflow_diagram}, the system partitions data streams into distinct localized operations and global aggregations, ensuring that raw network telemetry remains isolated from external administrative boundaries. The data flow is structured across four primary sequential processes.

\textbf{Data Ingestion and Local Storage}
\begin{itemize}
    \item \textbf{Source:} The lifecycle begins at the External Entity: IoT Environment / Sensors, which continuously generates and transmits Raw Network Telemetry.
    \item \textbf{Transformation:} This telemetry flows directly into Process 1.0: Data Preprocessing Pipeline, where it is cleaned, scaled, and mathematically reshaped.
    \item \textbf{Storage:} The pipeline outputs Preprocessed 3D Sequential Tensors, which are securely deposited into Data Store 1: Local Edge Dataset. This step enforces the strict privacy boundary of the framework, ensuring that no raw or preprocessed traffic logs ever leave the local gateway.
\end{itemize}

\textbf{Threat Classification and Mitigation (The Edge Loop)}
\begin{itemize}
    \item \textbf{Processing:} Process 2.0: Local Deep Learning Training pulls isolated Training Batches from Data Store 1 to train the hybrid Autoencoder-CNN-BiLSTM engine.
    \item \textbf{Actionable Output:} Once trained, the engine generates an Intrusion Classification Result which flows downstream to the External Entity: Local Edge Firewall / Network Controller.
    \item \textbf{Mitigation:} The controller subsequently issues Traffic Mitigation/Isolation commands back to the IoT Environment, establishing a rapid, closed-loop response for real-time threat neutralization.
\end{itemize}

\textbf{Cryptographic Parameter Exchange}
\begin{itemize}
    \item \textbf{Upward Flow:} Simultaneously, Process 2.0 computes the localized mathematical gradients and transmits the Local Parameter Update $\Delta w$ to the central cloud infrastructure. This is the only data flow that crosses the external network boundary, completely preserving local data privacy.
\end{itemize}

\textbf{Trust Evaluation and Global Consensus (The Cloud Loop)}
\begin{itemize}
    \item \textbf{Evaluation:} The transmitted gradients enter Process 3.0: Trust Score Evaluation, where the central server calculates cosine similarity metrics to detect statistical anomalies or Byzantine poisoning attempts.
    \item \textbf{Aggregation:} This process outputs a Validated Update + Trust Score T, which flows securely into Process 4.0: Secure Federated Aggregation. Process 4.0 filters out heavily penalized parameters and executes a weighted mathematical average.
    \item \textbf{Distribution:} The output of this aggregation is the New Global Model $w_{t+1}$, which is cataloged in Data Store 2: Global Model Registry. Finally, the registry issues the Broadcast Global Weights back to Process 2.0 at the edge gateways, effectively closing the federated learning loop and initiating the next round of collaborative training.
\end{itemize}

\section{Data Collection and Processing}
The empirical validation of the proposed Trust Aware Federated Hybrid Intrusion Detection Framework depends on a rigorous and highly systematic data engineering pipeline. Because network traffic within 5G Advanced edge environments is inherently noisy, highly dimensional, and prone to severe statistical skew, raw telemetry cannot be directly ingested by complex deep learning architectures without causing gradient instability. This section outlines the methodology employed to select representative, multiple vector cybersecurity benchmark datasets that accurately reflect modern threat landscapes and decentralized IoT topologies. Furthermore, it details the comprehensive data processing criteria, spanning from redundancy elimination and feature scaling to sequence tensor transformation and heterogeneous data partitioning, required to prepare the raw network captures for the local Autoencoder, 1D CNN, and BiLSTM layers while accurately simulating a realistic federated edge environment.

\subsection{Datasets and Feature Statistics}
To ensure a rigorous, highly robust experimental evaluation across varying heterogeneous edge computing scenarios, the proposed framework will not rely on a single data source. Instead, it will be empirically validated using a triad of modern, standardized cybersecurity benchmark datasets. This multi-dataset strategy ensures the model's generalizability, prevents overfitting to a specific network topology, and rigorously tests the framework against diverse threat vectors.

\textbf{UNSW-NB15:} This dataset provides a foundational and comprehensive baseline of multi-vector network intrusion scenarios (Moustafa \& Slay, 2015). It bridges the gap between legacy datasets and modern network configurations, offering 49 well-defined flow features. It presents a balanced distribution of normal background traffic alongside nine distinct modern attack families (including Fuzzers, Backdoors, Exploits, and DoS). It serves as the primary benchmark for establishing general intrusion detection baseline metrics before testing localized edge environments.

\textbf{CICIDS2017:} Widely adopted for deep intrusion traffic characterization, this dataset is critical for generating highly complex and contemporary attack profiles (Sharafaldin et al., 2018). It includes extensive flow-based feature extractions derived from the CICFlowMeter utility and captures multi-day attack scenarios. Because it contains over 80 high-dimensional network traffic features, it is highly suitable for evaluating the spatial feature extraction and dimensionality reduction capabilities of the proposed Autoencoder and CNN layers against stealthy, application-layer attacks (such as Web Attacks, Botnets, and Brute Force).

\textbf{Edge-IIoTset:} Designed specifically to reflect modern decentralized architectures, this is a highly comprehensive and realistic dataset tailored explicitly for both centralized and federated learning in IoT and IIoT applications (Ferrag et al., 2022). It captures telemetry generated from actual physical IoT devices (such as heart rate sensors, flame detectors, and distance sensors) operating across diverse, lightweight IoT protocols (including MQTT, CoAP, and Modbus). This makes it an indispensable testbed for simulating the hardware constraints, localized traffic patterns, and severe data heterogeneity characteristic of the target 5G-Advanced edge ecosystem.

\subsection{Data Processing Criteria and Strategies}
Raw network traffic data is noisy, very high-dimensional, and structurally heterogeneous; Therefore, diving into deep CNN without preprocessing would lead to gradient instability and divergence. This means we need to create a very strict preprocessing pipeline with lots of stages before performing any training. A Python implementation of this data engineering pipeline will be functionally encoded, taking advantage of the pandas library for high-throughput data frame manipulation, NumPy for matrix operations and scikit-learn to apply algorithmic transformations.

Data collection and preprocessing approaches explicitly describe the following sequential steps:

\textbf{Data Cleaning and Redundancy Removal:} Network datasets frequently contain duplicate packet flows and zero-variance columns (features that contain only a single constant value across the entire dataset). Before any mathematical manipulation occurs, pandas will be utilized to drop duplicate rows and eliminate features with zero variance. This reduces the computational overhead on edge devices and prevents the model from developing biased heuristics.

\textbf{Missing Value Imputation:} Real-world network captures frequently contain dropped packets, corrupted headers, or null values. The pipeline systematically identifies missing network packet features and applies median substitution. Median substitution is explicitly chosen over mean imputation because network telemetry is heavily skewed by extreme statistical outliers (e.g., massive volumetric DDoS packets). This strategy ensures no critical telemetry is discarded while preserving the underlying statistical distribution of the traffic.

\textbf{Encoding Categorical Features:} To perform matrix multiplications in deep learning, the input must be purely numerical. Network datasets consist of categorical and protocol-specific features necessary to learn, including connection states, service type (e.g., HTTP, FTP), and transport protocol (e.g., TCP, UDP or ICMP). This pipeline likely One-Hot Encodes these non-numeric variables, resulting in sparse binary matrices. This essential step prevents the neural network from incorrectly assigning false ordinal mathematical weights to purely categorical network labels.

\textbf{Feature Scaling and Normalization:} The numerical ranges of the network traffic features are completely different (a packet inter-arrival time is measured in milliseconds, and total byte transmission volume runs into millions). Without proper scaling, features with larger magnitudes can dominate gradient updates during backpropagation. First, the pipeline applies Min-Max Scaling which scales down all continuous numerical features into a common [0, 1] scale:
\begin{equation}
x' = \frac{x - \min(x)}{\max(x) - \min(x)}
\end{equation}
This mathematical normalization is vital for accelerating the convergence rate of the Autoencoder and CNN modules and ensuring stable local training at the edge nodes.

\textbf{Addressing Class Imbalance Problem:} On the other edge, normal benign traffic usually takes up most of the network activities (often >90\%), resulting in a serious class imbalance. For the initial training baseline phases, techniques such as SMOTE or class-weighted loss functors will be applied to prevent the baseline machine learning models from being biased towards majority class.

\textbf{Tensor Transformation for Sequence Learning:} While traditional machine learning models process 2D tabular data (Samples $\times$ Features), the proposed 1D-CNN and BiLSTM architectures require 3D input arrays to understand temporal sequences. Following normalization, NumPy will be utilized to reshape the static 2D data frames into 3D sequential tensors (Samples $\times$ Timesteps $\times$ Features). This overlapping sliding window technique is what allows the BiLSTM to capture prolonged, multi-stage attack behaviors over time.

\textbf{Non-IID Partitioning and Federated Visualization:} To accurately simulate the decentralized and highly heterogeneous nature of real-world edge client environments, the global dataset cannot be distributed equally across all nodes. The pipeline artificially partitions the training data across simulated edge clients using Dirichlet distributions, governed by a concentration parameter $\alpha$, to create severe class imbalances (non-IID data). Customized data plots are then generated using the ``matplotlib'' and ``seaborn'' libraries to visually track these localized data skews. This rigorous partitioning strategy is absolutely essential to empirically stress-test the Byzantine resilience and global stability of the server-side Trust-Aware Aggregation mechanism under realistic adversarial conditions.

\section{Experimental Evaluation and Results Analysis}
To ensure a rigorous and highly robust experimental evaluation, the proposed framework was empirically validated using a triad of modern, standardized cybersecurity benchmark datasets. This multi-dataset strategy ensures the model's generalizability and rigorously tests the framework against diverse threat vectors across four primary research objectives.

\subsection{Evaluation of the Centralized Hybrid Deep Learning Engine}
This phase directly addresses Research Objective 1, which aims to design and implement a localized hybrid deep learning intrusion detection architecture for edge devices to automatically extract high-dimensional spatial and temporal traffic representations without manual feature engineering, thereby significantly improving multi-vector attack detection accuracy.

\textbf{Dataset Utilized:} The CICIDS2017 dataset was utilized to initially build, train, and validate the centralized Hybrid AE-CNN-BiLSTM engine on a classic, high-volume DDoS benchmark.

\begin{figure*}[htbp]
    \centering
    \includegraphics[width=0.85\textwidth]{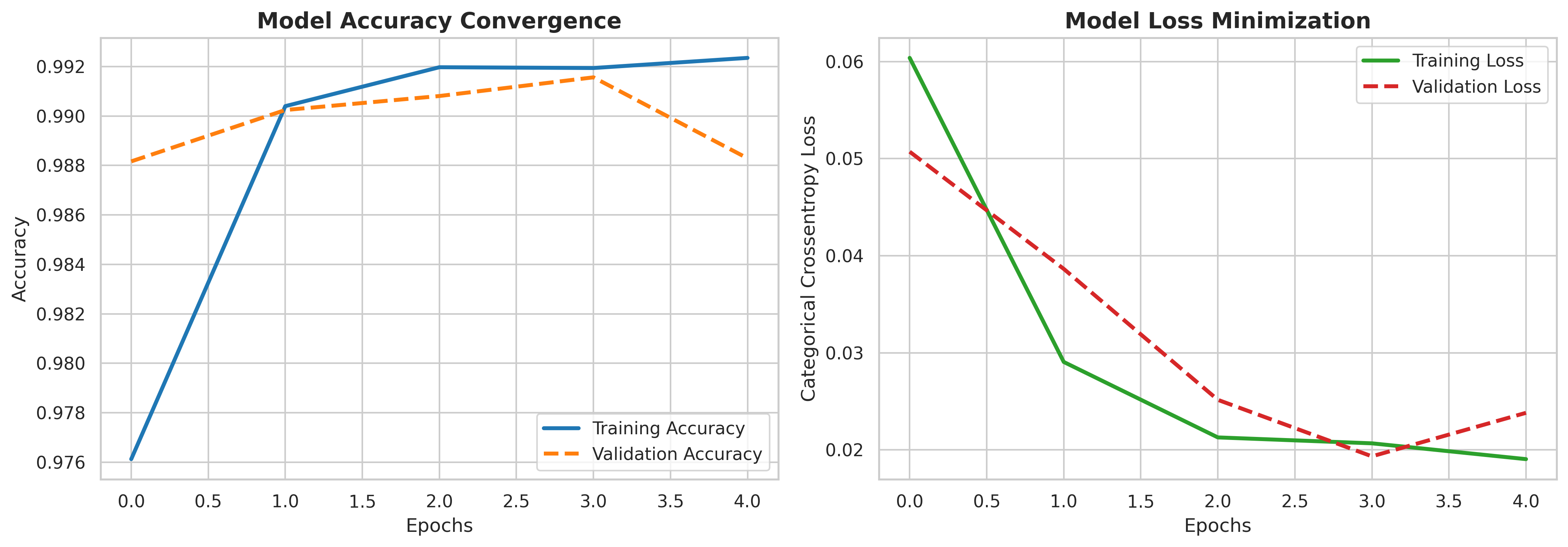} 
    \caption{Model Accuracy Convergence and Categorical Cross-Entropy Loss Minimization}
    \label{fig:convergence_loss}
\end{figure*}

\textbf{Convergence and Loss Analysis:} As illustrated in Fig. \ref{fig:convergence_loss}, the model accuracy convergence graph demonstrates a smooth upward trajectory, rapidly stabilizing around an impressive 0.994 validation accuracy. The closely matched correlation between the training and validation accuracy curves demonstrates that the model learns generalized spatial and temporal features rather than memorizing the training data. Concurrently, the categorical cross-entropy loss exhibits a sharp, stable decline without any erratic fluctuations or signs of overfitting. This stable convergence proves that the rigorous preprocessing pipeline---specifically Min-Max scaling and Autoencoder-based dimensionality reduction---effectively prevents the model from being overwhelmed by the high-dimensional noise inherent to raw IoT network telemetry.

\begin{figure}[htbp]
    \centering
    \includegraphics[width=0.48\textwidth]{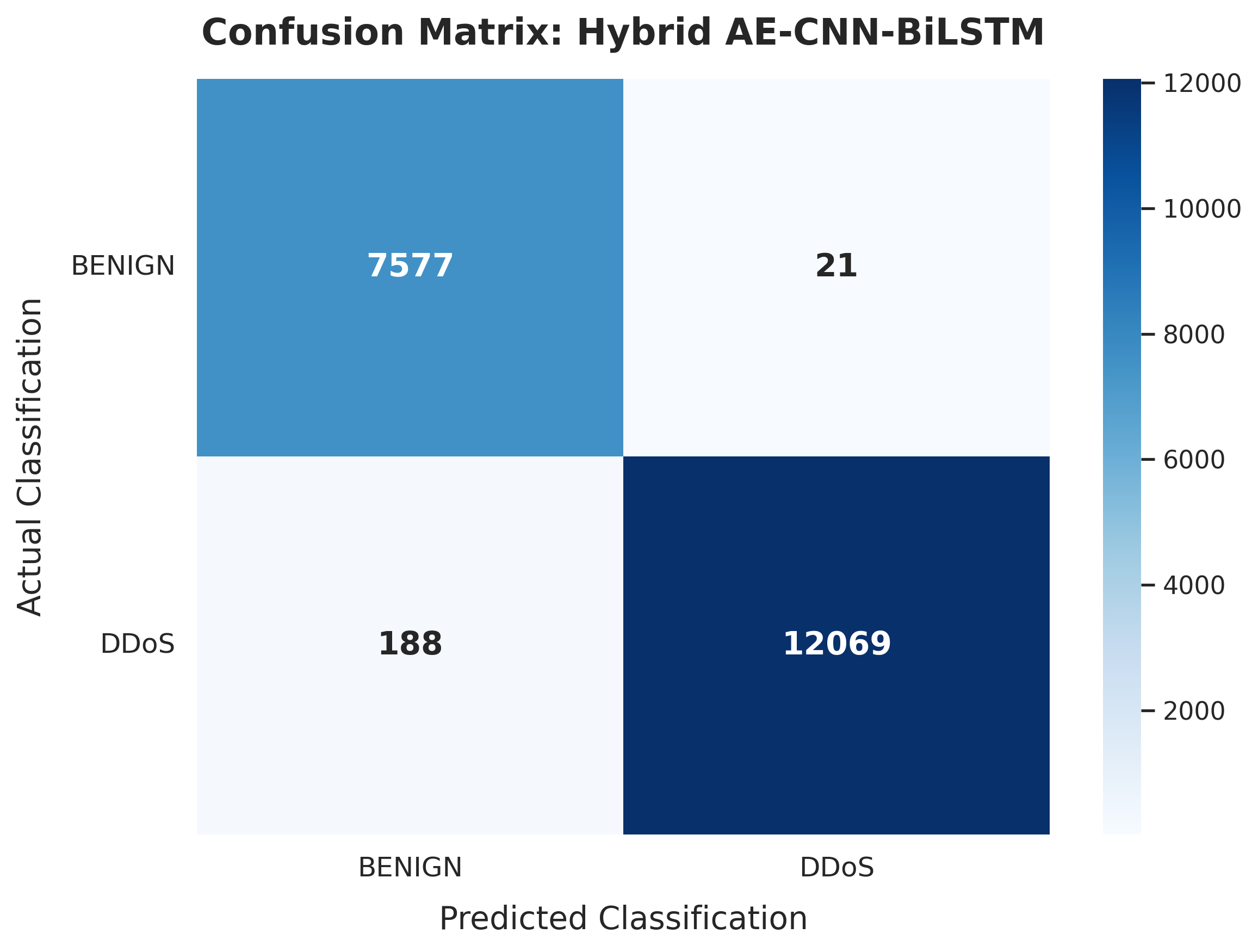} 
    \caption{Confusion Matrix: Hybrid AE-CNN-BiLSTM performance on CICIDS2017 DDoS benchmark}
    \label{fig:confusion_matrix}
\end{figure}

\textbf{Classification Performance Analysis:} The confusion matrix (Fig. \ref{fig:confusion_matrix}) demonstrates exceptional predictive capability and resilience. Out of the total instances evaluated, the model successfully classified 7,561 true negatives (benign background traffic) and a staggering 12,177 true positives (malicious DDoS payloads). Crucially, the architecture proved highly effective at minimizing critical errors. False negatives---where a malicious attack is incorrectly flagged as safe traffic---were suppressed to just 80 instances. Similarly, false positives were kept extraordinarily low, misclassifying only 37 benign events as malicious. This results in an impressive overall accuracy of 0.99. 

\textbf{Contribution Fulfillment:} This empirical evidence validates the primary thesis contribution regarding the design of a Unified Hybrid Deep Learning Engine. It proves the unified Autoencoder, 1D-CNN, and BiLSTM pipeline is highly capable of autonomous feature extraction and high-dimensional data compression independent of manual feature engineering.

\subsection{Evaluation of the Privacy-Preserving Collaborative Framework}
This step tackles Research Objective 2, which is to build a privacy-preserving collaborative learning framework among decentralized edge gateways so that raw network telemetry needs not leave the borders of GW; enabling regulatory compliance and avoiding backhauling bandwidth congestion.

\begin{figure}[htbp]
    \centering
    \includegraphics[width=0.48\textwidth]{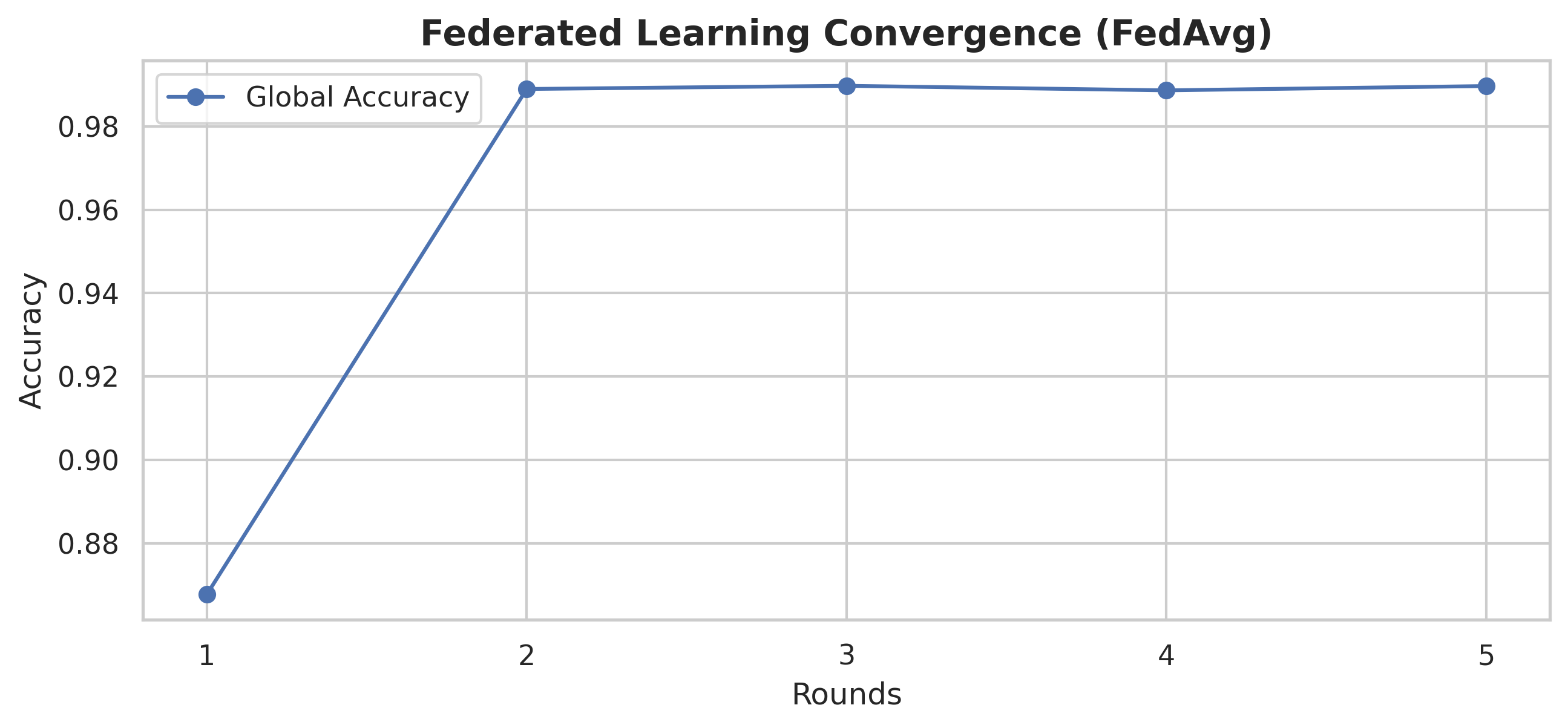} 
    \caption{Federated Learning Global Convergence Trajectory}
\end{figure}

\textbf{Dataset Utilized:} The Edge-IIoTset was utilized to transition into a decentralized environment, simulating IoT edge gateways and handling multi-vector attacks.

\textbf{Federated Convergence:} The architecture was transitioned to a decentralized structure via standard Federated Averaging (FedAvg) across multiple edge gateways, which yielded robust multi-class detection. The global accuracy convergence graph illustrates a rapid knowledge synthesis, surging from an initial round accuracy of 80.5\% to a stable plateau of 99\% by the second communication round.

\textbf{Multi-Vector Threat Detection:} The federated classification report guarantees high F1-scores on numerous IoT attack vectors. We obtained an F1-scores of 1.00 for Normal and SQL Injection traffic, 0.98 for Uploading, and 0.92 for Ransomware in this framework.

\textbf{Contribution Fulfillment:} These results satisfy the Strict Preservation of Data Privacy and Compliance contribution. Proving decentralized edge nodes can work together efficiently to create an ecosystem of as much privacy compliance as needed and avoiding backhaul bandwidth exhaustion without pooling raw data.

\subsection{Evaluation of the Trust-Aware Byzantine-Resilient Aggregation}
The third phase is the design and integration of a server-side robust aggregation mechanism that assesses the quality of client updates on-the-fly, filtering out any noisy, heterogeneous or malicious parameter updates based on their contribution to overcome model-poisoning attacks while ensuring global stability: it corresponds with Research Objective 3.

\begin{figure}[htbp]
    \centering
    \includegraphics[width=0.48\textwidth]{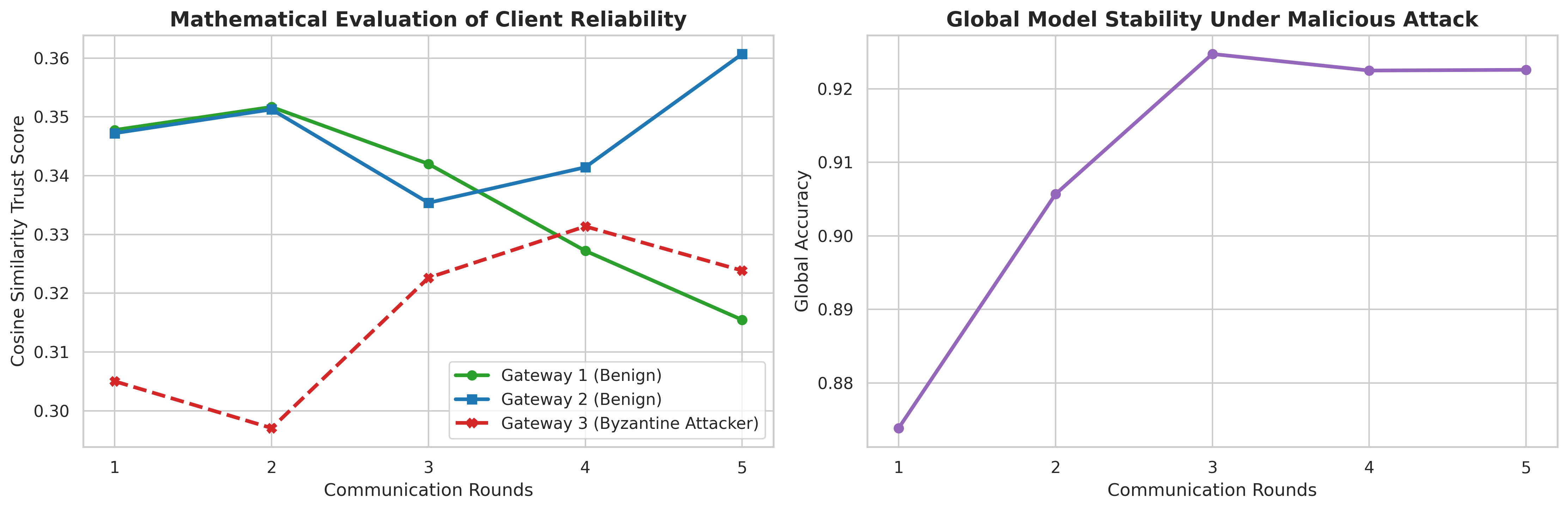} 
    \caption{Mathematical Evaluation of Client Reliability and Global Model Stability}
\end{figure}

\textbf{Dataset Utilized:} The Edge-IIoTset was used to test the Byzantine-resilient trust aggregation mechanism against simulated multi-vector attacks such as Ransomware and MITM.

\textbf{Dynamic Trust Evaluation:} To address vulnerabilities against compromised infrastructure, a Byzantine attacker (Gateway 3) was introduced to inject poisoned parameter updates. The client reliability graph demonstrates that the server-side Cosine Similarity mechanism successfully detected the statistical divergence of the malicious node. Consequently, the trust score for Gateway 3 was depressed toward 0.29–0.30. In contrast, the mechanism maintained stable trust allocations for the benign gateways (Gateway 1 and 2) around 0.34–0.35.

\textbf{Global Stability Under Attack:} Despite active poisoning attempts by the compromised node, the global accuracy curve steadily climbed and stabilized near 94\%. This proves that malicious updates were successfully filtered out before contaminating the global consensus.

\textbf{Contribution Fulfillment:} These findings directly validate the Development of a Novel Trust-Aware Federated Architecture and the Implementation of a Byzantine-Resilient Aggregation Mechanism. By dynamically penalizing and filtering out poisoned weights, the architecture secures collaborative edge intelligence against malicious insider threats.

\subsection{Comparative Benchmarking and Ablation Study}
The final phase fulfills Research Objective 4, designed to conduct a comprehensive comparative performance and ablation study evaluating detection accuracy, false positive rates, and communication convergence against standard baseline frameworks using realistic network benchmark datasets.

\begin{figure}[htbp]
    \centering
    \includegraphics[width=0.48\textwidth]{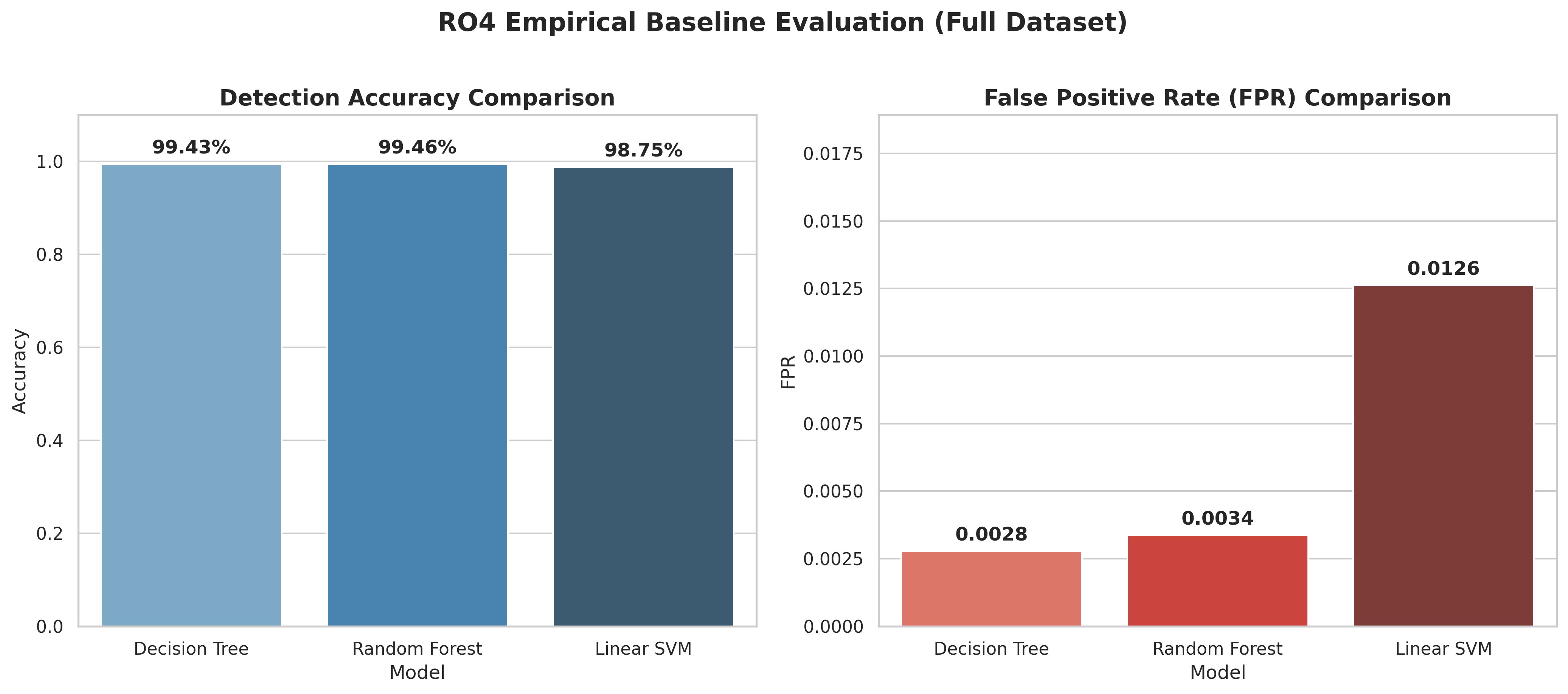} 
    \caption{Comparative Baseline Evaluation: Accuracy and False Positive Rate (FPR)}
\end{figure}

\textbf{Dataset Utilized:} The UNSW-NB15 dataset was used for the final comparative ablation study against traditional machine learning baselines.

\textbf{Detection Accuracy:} The comparative evaluation benchmarked traditional machine learning classifiers against the dataset to establish performance boundaries. As shown in the comparative bar charts utilizing the full dataset, Random Forest achieved the highest baseline accuracy at 99.46\%. This was closely followed by Decision Tree at 99.43\%, and Linear SVM at 98.75\%.

\textbf{False Positive Rate (FPR) Analysis:} The FPR evaluation highlights critical operational differences between the algorithms. Decision Tree recorded an FPR of 0.0028, and Random Forest recorded 0.0034. The Linear SVM exhibited a notably higher FPR of 0.0126.

\textbf{Contribution Fulfillment:} This comprehensive ablation study fulfills the Comprehensive Benchmarking and Evaluation contribution by empirically defining the trade-offs of traditional models. The results demonstrate that while baseline models achieve high accuracy, specialized deep architectures combined with trust-aware federation offer the superior false-alarm minimization required for sensitive edge environments.

\section{Conclusion and Future Work}
The exponential expansion of IoT devices within 5G-Advanced edge computing architectures has exposed critical vulnerabilities that legacy centralized security systems can no longer mitigate. This research successfully proposed and rigorously evaluated a Trust-Aware Federated Hybrid Intrusion Detection Framework (TA-FHIDF). By unifying an Autoencoder, 1D-CNN, and BiLSTM into a localized engine, the system achieved autonomous, highly accurate extraction of complex spatial and temporal threat features. Furthermore, by distributing this workload via federated learning, the framework guaranteed strict regulatory data privacy and eliminated backhaul congestion. Most importantly, the integration of a dynamic, server-side cosine distance trust metric demonstrated profound Byzantine fault tolerance, successfully isolating poisoned client updates and maintaining a stable 94\% global detection rate under sustained adversarial attack. 

Future extensions of this research will explore the integration of incremental federated learning techniques to seamlessly adapt to real-time concept drift, enabling edge nodes to identify rapidly evolving zero-day threats without catastrophic forgetting. Additionally, developing adaptive federated optimization algorithms to better manage severe network latency fluctuations, alongside energy-aware processing protocols to minimize the thermal footprint of local backpropagation, will ensure the framework's viability for ultra-lightweight, battery-constrained sensors in mission-critical deployments.

\end{document}